\documentclass{article}
\usepackage{spconf,amsmath,graphicx}
\usepackage{amssymb}
\usepackage{multirow}

\def\mathbi#1{\textbf{\em #1}}

\title{THE ICSTM+TUM+UP APPROACH TO THE 3RD CHIME CHALLENGE: SINGLE-CHANNEL LSTM SPEECH ENHANCEMENT WITH MULTI-CHANNEL CORRELATION SHAPING DEREVERBERATION AND LSTM LANGUAGE MODELS}
\name{Amr El-Desoky Mousa$^{1}$, Erik Marchi$^{2}$, Bj\"{o}rn Schuller$^{1,3}$}
\address{Author Affiliation(s)}

\address{ $^1$Chair of Complex \& Intelligent Systems, University of Passau, Passau, Germany \\
          \hspace{-15pt}$^2$Machine Intelligence \& Signal Processing group, MMK, Technische Universit\"{a}t M\"{u}nchen, Munich, Germany \\
          $^3$Department of Computing, Imperial College London, London, UK
}

\begin{document}
%\ninept
%
\maketitle
%

%%%%%%%%%%%%%%%%%%%%%%%%%%%%%%%%%%%%%%%%%%%%%%%%%%%%%%%%%%%%%%%%%%%%%%%%%%%%%%%%%%%%%%%%%%%%%%%%%%%%%%%%%%
\begin{abstract}

This paper presents our contribution to the 3rd CHiME Speech Separation and Recognition Challenge.
Our system uses Bidirectional Long Short-Term Memory (BLSTM) Recurrent Neural Networks (RNNs) for
Single-channel Speech Enhancement (SSE). Networks are trained to predict clean speech as well as 
noise features from noisy speech features. In addition, the system applies two methods of dereverberation
on the 6-channel recordings of the challenge. The first is the Phase-Error based Filtering (PEF) 
that uses time-varying phase-error filters based on estimated time-difference of arrival of the
speech source and the phases of the microphone signals. The second is the Correlation Shaping (CS)
that applies a reduction of the long-term correlation energy in reverberant speech. The Linear Prediction
(LP) residual is processed to suppress the long-term correlation. Furthermore, the system employs 
a LSTM Language Model (LM) to perform N-best rescoring of recognition hypotheses. Using the proposed
methods, an improved Word Error Rate (WER) of 24.38\% is achieved over the real eval test set. This 
is around 25\% relative improvement over the challenge baseline.

\end{abstract}
%%%%%%%%%%%%%%%%%%%%%%%%%%%%%%%%%%%%%%%%%%%%%%%%%%%%

\vspace{-2mm}
%%%%%%%%%%%%%%%%%%%%%%%%%%%%%%%%%%%%%%%%%%%%%%%%%%%%%%%%%%%%%%%%%%%%%%%%%%%%%%%%%%%%%%%%%%%%%%%%%%%%%%%%%%
\section{Introduction}
\label{sec:intro}

Automatic Speech Recognition (ASR) in real-world noisy and reverberated environments is a challenging
problem. The 3rd CHiME Speech Separation and Recognition Challenge (CHiME-3) addresses this problem 
by providing a testing platform for speech enhancement and recognition techniques \cite{Barker:2015}. 
The CHiME-3 scenario involves performing ASR for 6-channel microphone array data recorded via a 
multi-microphone tablet device being used in everyday, noisy environments. The data involves real
acoustic mixing in four various noise settings, namely caf\'{e}, street junction, public transport 
and pedestrian area. The recordings are divided into training, dev and eval sets. Each set features
different talkers and different instances of the same noise environment.

Methods for robust recognition of noisy speech can be categorized into two broad categories: the first
category involves {\bf front-end} enhancement of either the waveforms or the extracted features by 
removing noise and reverberation \cite{Virtanen:2012}. It is also possible to employ feature adaptations
to transform the corrupted features, or to use noise-robust features directly. The other category involves
improved {\bf back-end} ASR systems. One method is to adapt the models to noisy features, e.g., using 
multi-condition training or methods such as vector Taylor series \cite{Geiger:2014}. In particular, Deep
Neural Network (DNN) models have shown robust performance in recent years for both acoustic models
\cite{Hinton:2012} and language models \cite{naacl-hlt12:arisoy,desoky:2013,Mikolov:2012,Sundermeyer:2012}.

Neural networks for blind non-linear source separation have been extensively studied; e.g. in 
\cite{Karhunen:97,Tan:2001}. However, these works do not consider speech and noise model training.
Training of ASR feature enhancement models has been considered in \cite{Maas:2013}. Therein, RNN
Auto-Encoders (AE) are used to enhance cepstral-domain speech recognition features, but synthesis of 
time-domain signals is not considered. In \cite{Woellmer:2013, Weninger:2013}, a similar approach is % Weninger:2013
considered with LSTM architectures and is found superior to standard RNNs. In \cite{Lu:2013}, DNNs 
are used to map noisy to clean Mel-features, but the network outputs are synthesized directly into 
a time domain signal, instead of constructing a filter based on speech and noise magnitudes. In
\cite{Xia:2013}, a combination of unsupervised noise estimation and DNN based speech power spectrum
estimation is used to construct a Wiener filter. However, this work does not consider learning
noise models. In \cite{Narayanan:2013}, DNNs are considered to predict the ideal ratio mask in an
uncertainty decoding framework for ASR. However, the authors do not evaluate their models in terms
of separation quality. In this paper, we apply an approach that we previously described in 
\cite{Weninger:2014}. Thus, we use LSTM RNNs to model speech and noise features in a speech 
enhancement framework.

On the other side, reverberation severely degrades the performance of ASR. Suitable schemes for 
modeling reverberation are broadly applied such as the source-image method \cite{Allen:97,Peterson:86}.
Generally, a reverberant scenario consists of a source speech signal which propagates through an 
acoustic channel and is then captured by a microphone. The microphone signal, however, contains a 
reverberated version of the source signal. Thus, dereverberation techniques are applied on the 
microphone signal to output an estimate of the source signal \cite{Geiger:2014}. Many dereverberation 
algorithms have been developed over the last two decades \cite{Peterson:86}. Several strategies 
have been proposed, ranging from LP residual processing \cite{Yegnanarayana:2000} to multiple 
microphone array based techniques \cite{Griebel:99, ward:2001}. Further approaches have addressed 
blind system identification \cite{Xu:95} by using subspace decomposition \cite{Gannot:2003} and 
adaptive filters \cite{Huang:2005}. In this paper, we apply two multi-channel dereverberation 
techniques that we previously introduced in \cite{Geiger:2014}. The first technique is called 
Phase-Error based Filtering (PEF). It relies on time-delay estimation with time-frequency masking
\cite{Lai:2004, Aarabi:2004}. The second technique is called Correlation Shaping (CS) \cite{Gillespie:2003}.
It is based on LP and reduces the length of the equalized speaker-to-receiver impulse response.
Both techniques are applied on top of the enhanced recordings.

As a recognition back-end, our system relies on the baseline GMM/DNN acoustic models \cite{Barker:2015}.
However, we employ a state-of-the-art LSTM LM to rescore the N-best recognition hypotheses obtained 
using the provided standard trigram LM.

%This baseline system uses a standard trigram LM for recognition.
%As a step further, we employ a state-of-the-art LSTM LM to rescore the N-best recognition hypotheses.
 
%The DNN is trained using the standard procedure: pre-training
%using Restricted Boltzmann Machines (RBM), cross entropy training, and sequence discriminative 
%training using the state-level Minimum Bayes Risk (sMBR) criterion.

%To maintain compatibility with the 2nd CHiME challenge, the new challenge re-uses the WSJ evaluation framework. 
%Utterances are provided recorded in continuous audio with ground truth VAD annotations. 

\vspace{-2mm}
%%%%%%%%%%%%%%%%%%%%%%%%%%%%%%%%%%%%%%%%%%%%%%%%%%%%%%%%%%%%%%%%%%%%%%%%%%%%%%%%%%%%%%%%%%%%%%%%%%%%%%%%%%
\section{Evaluation database}
\label{sec:db}

The CHiME-3 scenario described in \cite{Barker:2015} involves ASR for a multi-microphone tablet device. 
4 various environments are selected: ca\'{f} (CAF), street junction (STR), public transport (BUS) and 
pedestrian area (PED). For each environment, two types of noisy speech data are provided, real and 
simulated. The real data consists of new 6-channel recordings of sentences from the WSJ0 corpus spoken 
in noisy environments. The simulated data is constructed by mixing clean utterances from that corpus 
into background recordings in the four noisy environments. For ASR evaluation, the data is divided into 
official training, dev and eval sets. %Details are provided in \cite{Barker:2015}.

\vspace{-2mm}
%%%%%%%%%%%%%%%%%%%%%%%%%%%%%%%%%%%%%%%%%%%%%%%%%%%%%%%%%%%%%%%%%%%%%%%%%%%%%%%%%%%%%%%%%%%%%%%%%%%%%%%%%%
\section{Methodology}
\label{sec:method}

%%%%%%%%%%%%%%%%%%%%%%%%%%%%%%%%%%%%%%%%%%%%%%%%%%%%
\subsection{Recurrent neural networks}
\label{ssec:rnn}

%%%%%%%%%%%%%%%%%%%%%%%%
\subsubsection{Standard RNN architectures}
\label{sssec:rnn}

The neural network architecture adopted for our Single-channel Speech Enhancement (SSE) model as well as our LM
is based on LSTM RNNs \cite{Graves:2013}. A RNN can be described as an automaton-like structure mapping from
a sequence of observations to a sequence of output features. These mappings are defined by activation weights 
and a non-linear activation function as in a standard Multi-Layer Perceptron (MLP). However, recurrent 
connections allow to access activations from past time. For an input sequence $x_1^T$, a RNN computes the
hidden sequence $h_1^T$ and the output sequence $y_1^T$ by performing the following operations for 
$t = 1 \hspace{3 pt} \text{to} \hspace{3 pt} T$:
{\small
\begin{eqnarray}
\label{eq:rnn}
  h_{t}  & = & \mathcal{H}(W_{xh} x_{t} + W_{hh} h_{t-1} + b_{h})\\ 
  y_{t}  & = & W_{hy} h_{t} + b_{y}
\end{eqnarray}}where $\mathcal{H}$ is the hidden layer activation function, $W_{xh}$ is the weight matrix 
between input and hidden layer, $W_{hh}$ is the recurrent weight matrix between hidden layer and itself, 
$W_{hy}$ is the weight matrix between the hidden and output layer, $b_{h}$ and $b_{y}$ are the hidden
and output layer bias vectors respectively.

In a standard RNN, $\mathcal{H}$ is usually an element-wise application of sigmoid function. Such a network
is usually trained using the Back-Propagation Through Time (BPTT) training \cite{Hinton:1986}, where a recurrent
network with $N$ timesteps is considered as an unfolded deep Feed-Forward Neural Network (FFNN) with $N$ hidden
layers and the error is propagated recursively from the hidden layer of the current timestep to the hidden layer
of the previous timestep. However, the error gradients can quickly vanish as they get propagated in time or in
rare cases grow exponentially \cite{Bengio:1994}. This is known as the vanishing gradient problem. One solution
to this problem is to consider only several steps of unfolding (truncated BPTT). 
%, which limits the memorization capability of the network.

\vspace{-3mm}
%%%%%%%%%%%%%%%%%%%%%%%%
\subsubsection{LSTM RNN architectures}
\label{sssec:lstm}

\begin{figure}[!htbp] %[htb]
  \centerline{\includegraphics[width=75mm]{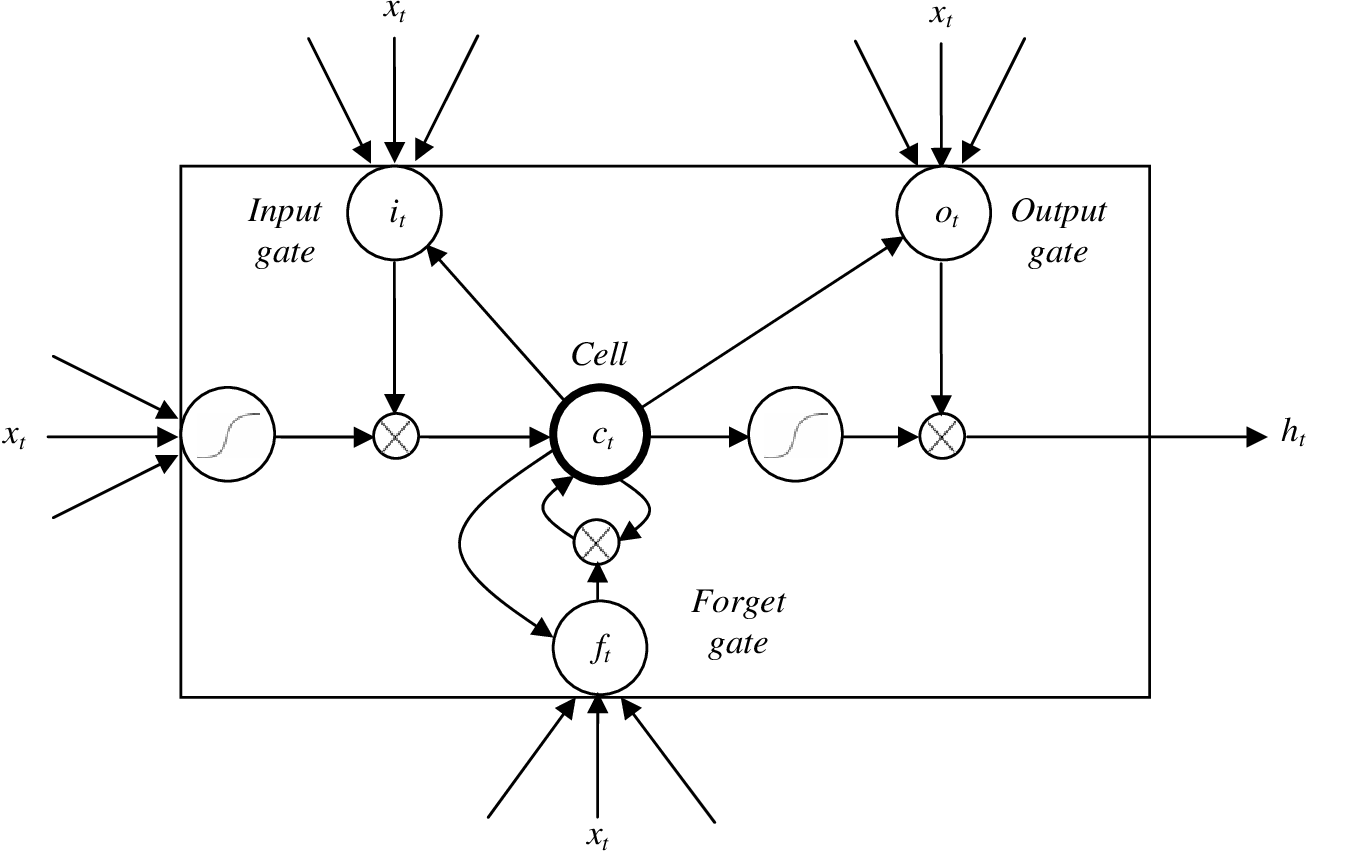}}
  \vspace{-6mm}
  \caption{Architecture of LSTM memory cell.}
\label{fig:memory-cell}
\vspace{-6mm}
\end{figure}

In \cite{Hochreiter:97}, an alternative RNN called Long Short-Term Memory (LSTM) RNN is introduced where the 
conventional neuron is replaced with a so-called {\it memory cell} that can be controlled by input, output 
and reset operations \cite{Gers:1999}. The purpose of this memory cell is to store information in such a way
that the corresponding gradient is properly scaled and never gets lost. This has been shown to overcome the
vanishing gradient problem of traditional RNNs \cite{Gers:2001}. Figure \ref{fig:memory-cell} illustrates the
architecture of a single LSTM memory cell. In this case, $\mathcal{H}$ can be described by the following composite
function:
{\small
\begin{eqnarray}
\label{eq:rnn}
  i_{t}  & = &  \sigma(W_{xi} x_{t} + W_{hi} h_{t-1} + W_{ci} c_{t-1} + b_{i})\\
  f_{t}  & = &  \sigma(W_{xf} x_{t} + W_{hf} h_{t-1} + W_{cf} c_{t-1} + b_{f})\\  
  c_{t}  & = &  f_{t} c_{t-1} \!+\! i_{t} \tanh(W_{xc} x_{t} \!+\! W_{hc} h_{t-1} \!+\! b_{c})\\ 
  o_{t}  & = &  \sigma(W_{xo} x_{t} + W_{ho} h_{t-1} + W_{co} c_{t} + b_{o})\\  
  h_{t}  & = &  o_{t} \tanh(c_{t})
\end{eqnarray}}where $\sigma$ is the sigmoid function. $i$,$f$,$o$ and $c$ are respectively the input, 
forget, output gates and cell activation vectors. %having the same size as the hidden vector $h$. 
The weight matrices from cell to gate vectors are diagonal \cite{Graves:2013}. 
%so that element $m$ in each gate vector receives input only from element $m$ of the cell vector \cite{Graves:2013}.

\vspace{-3mm}

%%%%%%%%%%%%%%%%%%%%%%%%%%%%%%%%%%%%%%%%%%%%%%%%%%%%
\subsubsection{Deep architectures}
\label{sssec:deep}

A major factor in the recent success of neural network models is the use of deep architectures obtained by stacking
multiple hidden layers on top of each other. In \cite{nips09:mohamed, Sainath:2012}, deep FFNNs have been successfully
used for acoustic modeling in ASR tasks. Also, in \cite{Graves:2013,Graves:2013b}, both deep conventional and LSTM RNNs
have been used for acoustic modeling in ASR. 
%Figure \ref{fig:deep-RNN} shows the general structure of a deep RNN with 
%$n$ hidden layers. If the conventional neuron of this figure is replaced with the memory cell architecture of
%Figure \ref{fig:memory-cell}, then a deep LSTM RNN is obtained.

%\begin{figure} [!htbp] %[htb]
%  \centerline{\includegraphics[width=35mm]{fig/deep_RNN}}
%  \vspace{-4mm}
%  \caption{Deep RNN architecture.}
%\label{fig:deep-RNN}
%%\vspace{-5mm}
%\end{figure}

%Unlike FFNNs, which typically use sliding windows of observations to provide context-sensitive, yet frame-by-frame
%predictions, RNNs also model dynamics of the output, which is arguably important for speech enhancement.  
Despite their context-sensitive nature, LSTM RNNs are well suited for online speech enhancement since they only require 
storing the current state of the automaton. In case that real-time capability is not needed, future context can also be 
exploited by adding a second set of layers which process the input feature sequences backwards, from $t = T$ to $t = 1$.
This extension leads to Bidirectional LSTM (BLSTM) RNNs. In a deep BLSTM RNN, activations from both directions are collected
in a single activation vector before passing them on as inputs to the next layer \cite{Graves:2013}.

\vspace{-2mm}
%%%%%%%%%%%%%%%%%%%%%%%%%%%%%%%%%%%%%%%%%%%%%%%%%%%%
\subsection{LSTM RNN language model}
\label{ssec:lstm-lm}

In a LSTM LM \cite{Sundermeyer:2012}, the timesteps correspond to the word positions in a training sentence. At every 
timestep, the network takes as input the word at the current position encoded using the $1$-of-$N$ binary encoding, 
where $N$ is the vocabulary size. An optional projection layer with linear (or sigmoidal) activation function is 
used to project the binary word vectors into vectors of real values representing points in continuous space. The 
projected vectors, or directly the binary vectors, are passed to one or more recurrent hidden layers with self connections
that implicitly take into account all the previous history words presented to the network. The output of the final 
hidden layer is passed to an output layer with a soft-max activation function to produce a correctly normalized probability
distribution. The target output at each word position is the next word in the sequence. The training is based on the 
cross entropy criterion which is equivalent to directly maximizing the likelihood of the training data, consequently 
minimizing the LM perplexity. Thus, for a given history $w_{1}^{n-1}$ the network can effectively predict the long-span 
conditional probability $p(w_n|w_{1}^{n-1})$ for any word $w_n$ of the vocabulary. 

\vspace{-2mm}
%%%%%%%%%%%%%%%%%%%%%%%%%%%%%%%%%%%%%%%%%%%%%%%%%%%%
\subsection{Single-channel speech enhancement}
\label{ssec:enhan}

Our SSE methodology is based on magnitude domain spectral subtraction. Let $\mathbf{X} \in \mathbb{R}^{F \times T}$
denote the magnitude spectrogram of a noisy speech signal with $F$ discrete Fourier frequency bins and $T$ observation frames.
From $\mathbf{X}$,a clean speech estimate $\mathbf{Y}$ is computed through:
{\small
\begin{equation}
\label{eq:filter-1}
\mathbf{Y} = \mathbf{X} \otimes (1 - \hat{\mathbf{N}}/\hat{\mathbf{X}})
\end{equation}}where $\otimes$ denotes element-wise multiplication and division is also element-wise. For traditional 
spectral subtraction, $\hat{\mathbf{X}} = \mathbf{X}$, so that the noise estimate is subtracted from the original noisy 
speech. Unsupervised estimation of $\hat{\mathbf{N}}$ is often done using minimum statistics \cite{Martin:2001}.

Our data-based speech enhancement algorithm uses an additional clean speech estimate $\hat{\mathbf{S}}$ in the above
filter, such that $\hat{\mathbf{X}} = \hat{\mathbf{S}} + \hat{\mathbf{N}}$. Thereby, the contribution of clean speech
and noise to the observed signal can be predicted. Popular models for speech include non-negative (sparse) coding by 
non-negative matrix factorization or Hidden Markov Models (HMMs). In our system, we use LSTM RNN based modeling via
supervised training of feature mappings similar to the denoising auto-encoder paradigm \cite{Xia:2013}. Due to their
recent success in noise robust ASR \cite{Woellmer:2013,Geiger:2013}, RNNs appear to be very well suited to capture the
dynamics of speech and noise as they directly model long-range context which cannot be approximated by `feature
frame stacking' in the general case.

In our system, networks are pre-trained to predict speech features from noisy speech features. Realistic instead of
white Gaussian noise is used for training like in \cite{Lu:2013}. Similarly, networks are trained to predict noise from
a convolutive mixture of speech and noise. During denoising, these estimates are used to construct a magnitude domain
filter as in Equation \ref{eq:filter-1}. As features for the neural networks, we use log Mel-scale spectrograms 
$\mathbf{X}' \in \mathbb{R}^{B \times T}$ with $B = 40$ frequency bands equally spaced on the Mel-frequency scale. Thus,
both amplitude and frequency are on a log-scale. These features have been proved highly successful for ASR with deep
LSTM RNNs \cite{Graves:2013}. Given predicted log Mel-features of speech and noise, $\hat{\mathbf{S}}'$ and $\hat{\mathbf{N}}'$,
the final filter equation is given by:
\vspace{-2mm}
{\small
\begin{equation}
\label{eq:filter-2}
\mathbf{Y} = \mathbf{X} \otimes \left[ 1 - \frac{\mathbf{M}^{-1}\exp(\hat{\mathbf{N}})}{\mathbf{M}^{-1} \left( \exp(\hat{\mathbf{S}}') + \exp(\hat{\mathbf{N}}')\right) } \right]
\vspace{-2mm}
\end{equation}}where $\mathbf{M}^{-1}$ denotes the `back-transformation' from Mel to magnitude spectra and exponentiation
is element-wise. Using Mel-spectra instead of magnitude or power spectra reduces the amount of speech features to be 
estimated. By reverting the Mel-scale transformation in the filter estimation, not in the estimated speech spectrogram, 
we avoid a loss of information due to the compression of the frequency axis. It is found that using the `ideal' filter
computed from `ground truth' speech and noise Mel-spectra provided perfect reconstruction in many cases. This enhancement
method is evaluated in our previous publication \cite{Weninger:2014}.

%A detailed description and evaluation of this methodology is found in our previous publication \cite{Weninger:2014}.

\vspace{-2mm}
%%%%%%%%%%%%%%%%%%%%%%%%%%%%%%%%%%%%%%%%%%%%%%%%%%%%
\subsection{Dereverberation}
\label{ssec:derev}

%Two multi-channel dereverberation methods are applied: Phase-Error based Filtering (PEF) and  Correlation Shaping (CS).
%A description of these methods is given in the following sections.

%The two methods are studied in our previous publication \cite{Geiger:2014}. A description is given in the following
%sections.

\vspace{-2mm}
%%%%%%%%%%%%%%%%%%%%%%%%%
\subsubsection{Phase-error based filtering}
\label{sssec:pef}

Phase-Error based Filtering (PEF) involves time-varying, or time-frequency, phase-error filters based on estimated 
Time-Difference of Arrival (TDOA) of the speech source and the phases of the microphone signals. The phase variance 
between two signals is defined as:
\vspace{-3mm}
{\small
\begin{equation}
\label{eq:pef-1}
  \psi_{\beta} =  \sum_{k=1}^{N} \sum_{\omega = -\omega_s}^{\omega_s} \theta_{\beta,k}^{2}(\omega),
\vspace{-5mm}
\end{equation}}where
{\small
\begin{equation}
  \theta_{\beta,k}(\omega) = \angle X_{1,k}(\omega) - \angle X_{2,k}(\omega) - \omega \beta
\label{eq:pef-2}
\end{equation}}indicates the level of noise and reverberation present in the speech signal. 
$\angle X_{1,k}$ and $\angle X_{2,k}$ are the phase spectra of the input signals at frame $k$, and
$\theta_{\beta,k}(\omega)$ is the minimized phase-error (PE) when $\beta$ equals the TDOA, $N$ indicates
the number of segments in the speech signal, and $\omega_s$ is the highest frequency of interest. 
The PE measures the time misalignment at each frequency bin. The overall PE can be reduced to:
{\small
\begin{equation}
  \theta_{\beta,k}(\omega) = \angle X_{1,k}(\omega) - \angle X_{2,k}(\omega)
\label{eq:pef-3}
\end{equation}}with the assumption that the input signals are time-aligned. The PE is used as a
reward-punish criteria to removing noise from multi-microphone speech signals. Time-frequency 
blocks with large PE are scaled down in amplitude, whereas, blocks with low PE are preserved. 
First, the PE is computed from the two phase spectra. Then, a masking function is applied as 
a weighting function for the amplitude spectrum of each channel. Spectra are later summed up
similarly as delay-and-sum. The parametrized scaling function:
{\small
\begin{equation}
  \eta(\omega) = \frac{1}{1 + \gamma^{\theta_{\beta,k}^{2}(\omega)}}
\label{eq:pef-3}
\end{equation}}is used in as a masking function to attenuate the time-frequency blocks, where
$\gamma$ is a fixed value. Higher values of $\gamma$ reduce high PE blocks prominently with 
a consequent improved performance in low Signal-to-Noise Ratio (SNR) scenarios and worse
performance in high SNR situations. PE based filtering is transferred to multi-microphone
signals by applying the parametrized scaling function on all possible pairs of microphones. Each
microphone pair $i$ and $j$ is processed by the masking function:
{\small
\begin{equation}
  \eta_{ij}(\omega) = \frac{1}{1 + \gamma^{\theta_{ij}^{2}(\omega)}}
\label{eq:pef-4}
\end{equation}}extended from Equation \ref{eq:pef-3}. Then, a modified geometric mean of the time-varying
functions \cite{Lai:2004} is used as follows:
{\small
\begin{equation}
  \Phi_i(\omega) = \left[ \prod_{j = 1,...,M \wedge i \neq j}^{} \eta_{ij}(\omega)  \right]^{\frac{1}{m}}
\label{eq:pef-3}
\end{equation}}where $M$ is the number of microphones and $m$ is a factor that affects the aggressiveness
of the algorithm. For a standard geometric mean, $m = M$. Using this approach, the estimation of high PE 
values is relevant in the mask averaging process. In fact, provided that a pair of microphones results in
a very high PE for a certain time-frequency block, the resulting scaling value will be close to zero. The
zero value is then kept in the geometrical averaging with the masking values for other pairs of microphones.
The enhanced spectrum $\hat{S}(\omega)$ is obtained by summing up the enhanced spectra processed by the 
multi-channel mask $\Phi_i(\omega)$, as follows:
\vspace{-2mm}
{\small
\begin{equation}
  \hat{S}(\omega) = \sum_{i=1}^{M} \Phi_i(\omega) X_i(\omega).
\vspace{-2mm}
\label{eq:pef-5}
\end{equation}}

A detailed description and evaluation of the PEF dereverberation method is found in our previous publication 
\cite{Geiger:2014}.

\vspace{-2mm}
%%%%%%%%%%%%%%%%%%%%%%%%%
\subsubsection{Correlation shaping}
\label{sssec:cs}

Correlation Shaping (CS) reduces the long-term correlation in the LP residual of reverberant speech. This approach improves both the 
audible quality and ASR accuracy of reverberant speech \cite{Gillespie:2003}. CS modifies the correlation structure
of the processed speech signal $y$. Assuming that an array of $M$ microphones records a speech source, the signal
observed by the $m$th microphone $x_m$ is processed by an adaptive linear filter $g_m$ in order to minimize the 
weighted Mean Square Error (MSE) between the actual output autocorrelation sequence $\mathbi{R}_{yy}$, and the 
desired output autocorrelation sequence $\mathbi{R}_{dd}$. The adaptive linear filters are continuously adjusted
via a set of feedback functions in order to minimize the MSE. Gradient Descent (GD) is used to perform the minimization.
The gradient relies on the output autocorrelation $\mathbi{R}_{yy}$, the cross-correlation between the output and 
input, $\mathbi{R}_{yx_m}$, and the desired output autocorrelation $R_{dd}$. The autocorrelation sequence 
$R_{x_{m}x_{m}}(\tau)$ of the multi-channel input sequence $x_m(n)$ is given by:
\vspace{-2mm}
{\small
\begin{equation}
  R_{x_{m}x_{m}}(\tau) = \sum_{n=0}^{N-1} x_m(n) x_m(n-\tau).
\vspace{-2mm}
\label{eq:cs-1}
\end{equation}}

CS is defined as a multi-input single-output linear filter:
\vspace{-2mm}
{\small
\begin{equation}
  y(n) = \sum_{m=0}^{M-1} \mathbi{g}_m^T(n) \mathbi{x}_m(n).
\vspace{-2mm}
\label{eq:cs-2}
\end{equation}}

The autocorrelation sequence $R_{yy}(\tau)$ of the output signal $y(n)$ is expressed as follows:
\vspace{-2mm}
{\small
\begin{equation}
  R_{yy}(\tau) = \sum_{n=0}^{N-1} y(n) y(n-\tau).
\vspace{-2mm}
\label{eq:cs-3}
\end{equation}}where $N$ is the number of samples over which autocorrelation is computed, $\tau$ is the correlation 
lag. The scope of CS is to minimize the weighted MSE given by:
{\small
\begin{equation}
  e(\tau) = W(\tau) \left( R_{yy}(\tau) - R_{dd}(\tau)  \right)^2, 
\label{eq:cs-4}
\end{equation}}where $W(\tau)$ is a real value weight. The larger $W(\tau)$ is, the more relevant the error at a specific
lag is. For dereverberation purposes, the LP residual is fed into the correlation shaping processor, and the target output
correlation is set to be $R_{dd}(\tau) = \delta(\tau)$. By further exploiting the autocorrelation symmetry, the gradient
can be simplified as:
{\small
\begin{equation}
  \nabla_m(l) \!=\! \sum_{\tau > 0} W(\tau) R_{yy}(\tau) \left( R_{yx_m}(l \!-\! \tau) + R_{yx_m}(l \!+\! \tau)  \right).
\vspace{-2mm}
\label{eq:cs-5}
\end{equation}}

This gradient is used in the filter update equation:
{\small
\begin{equation}
  g_m(l , n + 1) = g_m(l , n) - \mu \nabla_{m}^{'}(l),
\label{eq:cs-6}
\end{equation}}where $\mu$ is the learning rate parameter and $\nabla_{m}^{'}(l)$ is given by:
{\small
\begin{equation}
  \nabla_{m}^{'}(l) = \frac{\nabla_{m}(l)}{\sqrt{\sum_{m}\sum{l}\nabla_m^2(l)}}
\vspace{-2mm}
\label{eq:cs-7}
\end{equation}}

The dereverberated speech signal is obtained by applying the equalizer $g(l,n)$ to the input signal. Considering that 
the reverberation time affects significantly audio quality and ASR accuracy, a `don't care' region is introduced and 
applied to autocorrelation lags closed to the zeroth lag in order to improve the suppression of long-term components.
This region modifies the gradient in Equation \ref{eq:cs-5} and controls the value of the first autocorrelation lag.
Details about the CS dereverberation method can be found in our previous publication \cite{Geiger:2014}.

\vspace{-2mm}
%%%%%%%%%%%%%%%%%%%%%%%%%%%%%%%%%%%%%%%%%%%%%%%%%%%%%%%%%%%%%%%%%%%%%%%%%%%%%%%%%%%%%%%%%%%%%%%%%%%%%%%%%%
\section{System configuration}
\label{sec:pagestyle}

%%%%%%%%%%%%%%%%%%%%%%%%%%%%%%%%%%%%%%%%%%%%%%%%%%%%
\paragraph*{ASR Baseline:}
Our system makes use of the challenge ASR baseline \cite{Barker:2015}, which is a state-of-the art GMM/DNN system based 
on Kaldi toolkit \cite{Povey:ASRU2011}. The {\it GMM sub-system} uses 13-dimensional MFCC feature vectors. LDA is used 
to project a concatenation of 7 consecutive vectors in a sliding window (91 components) to 40 components. The system 
uses HMMs with 2500 tied triphone states modeled by 15,000 Gaussians. Maximum Likelihood Linear Transformation (MLLT), 
and feature space Maximum Likelihood Linear Regression (fMLLR) with Speaker Adaptive Training (SAT) are applied. The 
{\it DNN sub-system} uses 40-dimensional log Mel-filterbank features instead of MFCCs \cite{Hinton:2012, Graves:2013}. 
It has 7 layers with 2048 neurons each. The input layer has 440 units (5 frames of left and right context). The DNN is 
trained using the standard procedure: pre-training using Restricted Boltzmann Machines (RBMs), cross entropy training, 
and sequence discriminative training using the state-level Minimum Bayes Risk (sMBR) criterion. In addition, a baseline
signal enhancement is provided, which transforms the multi-channel noisy signal into a single-channel enhanced signal 
suitable for ASR \cite{Barker:2015}.

\vspace{-3mm}
%%%%%%%%%%%%%%%%%%%%%%%%%%%%%%%%%%%%%%%%%%%%%%%%%%%%
\paragraph*{Single-channel enhancement:} % To be modified After skyping with Erik

We use BLSTMs for prediction of either speech or noise features with 3 hidden layers. BLSTMs have 128 units per direction.
Feed-forward layers with 64 units are inserted after each BLSTM layer in order to reduce the number of parameters. Networks
are trained exactly as described in \cite{Weninger:2014} using the BPTT training algorithm. %For speech feature prediction, we use reverberated, yet noise-free features as training targets.
To prevent over-fitting, Gaussian noise with zero mean and 0.1 standard deviation is added to the inputs. Input and
target features are standardized to zero mean and unit variance, and delta regression coefficients of the feature contours 
are added. The sum-squared errors at the output layer per sequence is used as a cost function. 

%We use BLSTMs for prediction of either speech or noise features with 3 hidden layers. BLSTMs have 128 units per direction.
%Feed-forward layers with 64 units are inserted after each BLSTM layer in order to reduce the number of parameters. Networks
%are trained on the noisy and reverberated Audio-Visual Interest Corpus (AVIC) training set \cite{Schuller:09-BBR} exactly 
%as described in \cite{Weninger:2014}. For speech feature prediction, we use reverberated, yet noise-free features as training
%targets. To prevent over-fitting, Gaussian noise with zero mean and 0.1 standard deviation is added to the inputs. Input and
%target features are standardized to zero mean and unit variance, and delta regression coefficients of the feature contours 
%are added. The sum-squared errors at the output layer per sequence is used as a cost function. 

%the validation set error is 
%evaluated after each training epoch and training is aborted once the validation set error has converged. Additionally,
%sequences are shuffled in random order.

%on the noisy training set. Features from the close-talking microphone (considered as clean speech) are used as training targets.

\vspace{-3mm}
%%%%%%%%%%%%%%%%%%%%%%%%%%%%%%%%%%%%%%%%%%%%%%%%%%%%
\paragraph*{Dereverberation:}
PEF and CS are performed on top of the SSE. PEF is evaluated using a frame size of 1024 samples. Smaller frame sizes
result in less reliable phase estimates causing artifacts and distortions in the reconstructed signal. A frame shift
of 10ms is applied. $\gamma$ was set to 0.01 in order to avoid an aggressive masking that is suitable only in low SNR
conditions. $m$ was set to $M$ in order to obtain the geometric mean of the signal and avoid severe speech distortions.
CS is performed by estimating autocorrelation functions on the whole speech segment. We applied 62.5ms long equalizers,
a 18.7ms long `don’t care' region and exponential weighting. CS is performed up to $\tau_{max} =$ 62.5ms.

\vspace{-3mm}
%%%%%%%%%%%%%%%%%%%%%%%%%%%%%%%%%%%%%%%%%%%%%%%%%%%%
\paragraph*{LSTM LM:}
The standard LM for CHiME-3 is a trigram backoff LM with 5k vocabulary trained on the official LM training data, which consists 
of around 1.6M sentences (40.5M running words). Training a LSTM LM on this complete data requires impractically huge resources.
Therefore, a fraction of the training sentences is selected based on a sentence-level relevance measure with respect to the 
development set. Thus, a 5-gram backoff LM, called {\it dev-LM}, is estimated from the development text, then all the sentences
of the LM training data are ranked based on their perplexity with this {\it dev-LM}. A top most 80k sentences are selected with
around 2M running words, which is only 5\% of the total training data. This procedure allows for estimating the most relevant part
of the training text based on domain similarity with the development set. Our LSTM network uses one hidden layer of 300 units. 
Both input and output layers have 5k units (similar to the vocabulary size). No projection layer is used after the input layer.
The long-span probabilities of the LSTM LM are linearly interpolated with a background 5-gram backoff LM trained on the complete 
training data. The interpolation weight is optimized on the development set. The backoff LMs are estimated with Modified 
Kneser-Ney (MKN) smoothing using the SRILM toolkit \cite{icslp02:Stolcke}.

% SUMMARY of PPLs:
%ppl/dt05.text.gz.mylm-3.ppl:0 zeroprobs, logprob= -13206.5 ppl= 59.9701 ppl1= 76.4564
%ppl/dt05.text.gz.mylm-4.ppl:0 zeroprobs, logprob= -12737.3 ppl= 51.8519 ppl1= 65.5385
%ppl/dt05.text.gz.mylm-5.ppl:0 zeroprobs, logprob= -12631.7 ppl= 50.1823 ppl1= 63.3051
%ppl/dt05.text.gz.tglm.ppl:0 zeroprobs, logprob= -13356.3 ppl= 62.8213 ppl1= 80.3125
%ppl/et05.text.gz.mylm-3.ppl:0 zeroprobs, logprob= -9707.81 ppl= 58.5201 ppl1= 74.7086
%ppl/et05.text.gz.mylm-4.ppl:0 zeroprobs, logprob= -9285.9 ppl= 49.0341 ppl1= 61.9375
%ppl/et05.text.gz.mylm-5.ppl:0 zeroprobs, logprob= -9188.87 ppl= 47.0796 ppl1= 59.3237
%ppl/et05.text.gz.tglm.ppl:0 zeroprobs, logprob= -9841.17 ppl= 61.8848 ppl1= 79.2696

\vspace{-3mm}
%%%%%%%%%%%%%%%%%%%%%%%%%%%%%%%%%%%%%%%%%%%%%%%%%%%%
\paragraph*{LSTM Toolkit:}
In all our experiments, LSTMs are trained and evaluated using our own open-source implementation named CURRENNT
(CUDA RecuRrEnt Neural Network Toolkit)\footnote{http://sourceforge.net/p/currennt} \cite{Weninger:currennt:2014}.
CURRENNT uses Graphical Processing Units (GPUs) to speed up computation. Since in the case of RNNs, parallelization 
cannot be performed across timesteps due to the temporal dependencies, it parallelizes computations across sequences, 
for each timestep. This leads to a `semi-online' GD algorithm, where the weights of the network are updated after 
each batch of parallel sequences. 

\vspace{-3mm}
%%%%%%%%%%%%%%%%%%%%%%%%%%%%%%%%%%%%%%%%%%%%%%%%%%%%%%%%%%%%%%%%%%%%%%%%%%%%%%%%%%%%%%%%%%%%%%%%%%%%%%%%%%
\section{Results}
\label{sec:res}

Table \ref{tab:res} shows the WER performance of the GMM and DNN systems for different test sets and training
conditions. For a GMM system trained on clean data (original WSJ0 recordings), recognition of noisy test speech
shows the poorest performance. However, in this case, the simulated data can be used to approximately predict 
the performance of the real data. For a baseline enhanced test speech, the performance improved significantly. 
Yet, a big difference appears between the performance of the real and simulated data; this is most probably due
to the limitations of both the acoustic simulation baseline and the enhancement baseline. For SSE of channel 5,
performance improvement is only seen on the real eval test set. However, for other test sets, noticeable performance
degradations are recorded. This is due to the potential mismatch between the originally clean training conditions
and the enhanced test data, as well as the limitation of single channel processing. Applying additional PEF and CS
dereverberation to the SSE enhanced test sets leads to significant performance improvements for real data with 
almost comparable values.

For a GMM system trained on noisy data (non-enhanced channel 5), the performances are comparable after applying each
enhancement method to the test sets. In addition, performance similarity is almost kept for real and simulated data. 
%The best WER achieved for the real dev test set is 16.51\% after applying PEF dereverberation to the SSE enhanced signal.
%Whereas, the best WER achieved for real eval test set is 27.74\% after applying the CS dereverberation.

For a DNN system trained on baseline enhanced data, the best performance is recorded for the baseline enhanced test sets.
This is essentially expected due to the direct match between the training and testing conditions. Nevertheless, it is worth
observing that, among the other enhancement methods, the CS approach achieves the best performance. Therefore, we train a full
DNN system on the CS dereverberated training data. This achieves the best results so far on both real dev and eval test sets.
In addition, the simulated data shows a good capability to predict the performance of the real data.

Using a DNN system with an additional N-best rescoring via a LSTM LM interpolated with a background
5-gram backoff LM leads to further improvements in WERs. The N-best rescoring adds around 5\% relative WER improvement
for the case of baseline enhanced train and test data; and around 4\% WER improvement for the case of CS dereverberated 
train and test data compared to the non-rescored experiments. The best overall results are achieved using a DNN system 
with additional N-best rescoring via LSTM LM. Around 25\% relative WER improvement is achieved compared to the official challenge baseline results (given by line 11 
in Table \ref{tab:res}). Table \ref{tab:best-sys} shows the detailed WERs for our best system for every environment and 
every dev/eval test set.

%In case of the baseline enhanced train and test data, applying an additional N-best rescoring via a LSTM LM interpolated with 
%a background 5-gram backoff LM leads to considerable improvements in WERs. 

%However, still the best performance is observed in the case of 
%dereverberated test sets tested with a GMM system trained with noisy data. Table \ref{tab:best-sys} shows the detailed WERs
%for the best system for every environment and every dev/eval test set.

\begin{table}[!htbp] %[t,h]
%\vspace{2mm}
\caption{\label{tab:res} WER performance of the GMM and DNN systems for real and simulated, dev and eval test sets.
Models are trained/tested on: clean, noisy or enhanced noisy data. %The following enhancement methods are examined:
{\bf enhan}: baseline enhancement (all channels); 
{\bf SSE5}: single-channel speech enhancement (channel 5); 
{\bf PEF}: dereverberation via phase-error based filtering (all channels);
{\bf CS}: dereverberation via correlation shaping (all channels).
{\bf DNN+}: DNN acoustic model + N-best rescoring via LSTM LM.
}
\vspace{2mm}
\centerline{
\begin
{tabular}{|c|c|c|c|c|c|c|}
\hline
                               &                               &                    & \multicolumn{2}{c|}{\bf dev test}  & \multicolumn{2}{c|}{\bf eval test} \\
\cline{4-7}
\bf model                      & \bf train                     & \bf test           & \bf real  & \bf sim  & \bf real & \bf sim  \\
\hline
\hline
\multirow{10}{*}{\small GMM}   & \multirow{5}{*}{\small clean} & {\small noisy}     & 54.45     & 50.56    & 79.01    & 63.83 \\
                               &                               & {\small enhan}     & 41.69     & 21.88    & 75.86    & 25.86 \\
                               &                               & {\small SSE5}      & 41.83     & 39.12    & 66.84    & 46.96 \\
                               &                               & {\small PEF}       & 33.91     & 31.94    & 57.67    & 40.92 \\
                               &                               & {\small CS}        & 35.27     & 37.84    & 56.27    & 46.02 \\
\cline{2-7}
                               & \multirow{5}{*}{\small noisy} & {\small noisy}     & 18.71     & 18.82    & 33.95    & 21.92 \\
                               &                               & {\small enhan}     & 18.59     & 10.06    & 32.11    & 11.30 \\
                               &                               & {\small SSE5}      & 21.12     & 20.16    & 36.10    & 21.50 \\
                               &                               & {\small PEF}       & 16.51     & 16.26    & 28.79    & 19.55 \\
                               &                               & {\small CS}        & 17.03     & 18.81    & 27.74    & 20.92 \\
\hline
\multirow{5}{*}{\small DNN}    & \multirow{4}{*}{\small enhan} & {\small enhan}     & 17.76     & 8.67     & 32.71    & 10.79 \\
                               &                               & {\small SSE5}      & 33.67     & 29.63    & 55.96    & 35.75 \\
                               &                               & {\small PEF}       & 25.95     & 30.47    & 40.76    & 42.98 \\
                               &                               & {\small CS}        & 23.49     & 29.87    & 36.51    & 40.77 \\
\cline{2-7}
                               & {\small CS}                   & {\small CS}        & 14.08     & 15.34    & 25.53    & 22.32 \\
\hline
\multirow{2}{*}{\small DNN+}   & {\small enhan}                & {\small enhan}     & 16.71     & 7.82     & 30.81    & 10.11 \\
\cline{2-7}
                               & {\small CS}                   & {\small CS}        &\bf 13.71  &\bf 14.56 &\bf 24.38 &\bf 21.45   \\
\hline
\end{tabular}}
\end{table}

%%%%%%% BEST SYSTEM DETAILS

\begin{table}[!htbp] %[t,h]
%\vspace{2mm}
\caption{\label{tab:best-sys} Detailed WER performance for the best system with the least WER on real dev test set (row 15 of Table \ref{tab:res}). 
}
\vspace{2mm}
\centerline{
\begin
{tabular}{|c|c|c|c|c|}
\hline
\multirow{2}{*}{\bf environment} & \multicolumn{2}{c|}{\bf dev test}  & \multicolumn{2}{c|}{\bf eval test} \\
\cline{2-5}
                                 &  \bf real   & \bf sim   & \bf real & \bf sim  \\
\hline
\hline
BUS                              & 15.33  & 11.71  & 30.68  & 17.20    \\
CAF                              & 14.48  & 19.51  & 27.31  & 24.24    \\
PED                              & 9.96   & 12.08  & 21.26  & 22.38    \\
STR                              & 15.07  & 14.94  & 18.29  & 21.98    \\
\hline
\end{tabular}}
\end{table}

%\begin{table}[!htbp] %[t,h]
%%\vspace{2mm}
%\caption{\label{tab:best-sys} Detailed WER performance for the best system with the least WER on real eval test set (row 10 of Table \ref{tab:res}).
%}
%\vspace{2mm}
%\centerline{
%\begin
%{tabular}{|c|c|c|c|c|}
%\hline
%\multirow{2}{*}{\bf environment} & \multicolumn{2}{c|}{\bf dev test}  & \multicolumn{2}{c|}{\bf eval test} \\
%\cline{2-5}
%                                 &  \bf real   & \bf sim   & \bf real & \bf sim  \\
%\hline
%\hline
%BUS                              & 19.21  & 14.44  & 36.79  & 16.59   \\
%CAF                              & 17.68  & 25.47  & 30.30  & 23.91   \\
%PED                              & 13.85  & 16.24  & 24.25  & 20.96   \\
%STR                              & 17.39  & 19.09  & 19.63  & 22.23   \\
%\hline
%\end{tabular}}
%\end{table}

%%%%%%%%%%%%%%%%%%%%%%%%%%%%%%%%%%%%%%%%%%%%%%%%%%%%%%%%%%%%%%%%%%%%%%%%%%%%%%%%%%%%%%%%%%%%%%%%%%%%%%%%%%
\section{Conclusions}
\label{sec:conc}

This paper has presented our approach to the 3rd CHiME Separation and Recognition Challenge (CHiME-3),
employing single-channel speech enhancement via a fully data-based paradigm using deep BLSTM RNNs. In addition, 
two multi-channel dereverberation methods called phase-error based filtering and correlation shaping have been
applied using the 6-channel audio recordings of the challenge to estimate and filter the reverberation. 
Furthermore, an improved LM based on LSTM RNNs has been utilized to perform N-best rescoring of the recognition
hypotheses. Experiments have been performed according to the official challenge guidelines. %with the provided datasets in addition to some external data sources. 

The correlation shaping dereverberation has led to slightly better performance compared to the phase-error based
filtering approach. This agrees with the common wisdom that reducing the length of the equalized speaker-to-receiver
impulse response can improve the audible quality and ASR accuracy. Both methods have achieved better performance
in comparison to the single-channel enhancement alone. Moreover, the application of N-best rescoring via the LSTM
LM has led to improved performance.

%better performances compared to the single-channel enhancement alone. 
%with comparable performances to the phase-error based filtering. This agrees with the common wisdom that reducing the length of the equalized 
%speaker-to-receiver impulse response can improve the audible quality and ASR accuracy. Moreover, the application 
%of N-best rescoring via the LSTM LM has led to improved performances.

Further improvements are possible by improving the back-end recognition using, for example, BLSTM based 
acoustic models, and discriminatively trained GMMs. %In addition, the LSTM LM rescoring is not yet applied to the hypotheses of the best experiment. 
%In addition, the BLSTMs of the single-channel speech enhancement need to be retrained on the official training set only. 
In addition, the LSTM LM needs to be trained on more data from the official dataset.
It is also worth trial to train two full DNN systems using the single-channel enhanced and dereverberated
data via phase-error based filtering in order to draw a fair comparison among the employed enhancement techniques.

%%%%%%%%%%%%%%%%%%%%%%%%%%%%%%%%%%%%%%%%%%%%%%%%%%%%%%%%%%%%%%%%%%%%%%%%%%%%%%%%%%%%%%%%%%%%%%%%%%%%%%%%%%
\section{Acknowledgements}
\label{sec:foot}

The research leading to these results has received funding from the European Union’s Horizon 2020 Programme through the
Research Innovation Action No. 645378 (ARIA-VALUSPA); and from the German Federal Ministry of Education, Science, Research
and Technology (BMBF) under grant agreement No. 16SV7213 (EmotAsS).

% start a separate page for references
\newpage
\bibliographystyle{IEEEbib}
\bibliography{references}
%\bibliography{strings,refs}

\end{document}